\documentclass[nofootinbib,prd, preprint]{revtex4-1}
\usepackage{graphicx}
\usepackage{longtable}
\usepackage{amsmath}
\usepackage{amssymb}
\usepackage{subfigure}
\usepackage{microtype}
\usepackage{hyperref}
\usepackage{stackengine}

\usepackage{color}

\newcommand{\mpl}{{M_{\rm {pl}}}}

\newcommand{\dd}{\, {\rm d}}
\newcommand{\gsim}{\;\mbox{\raisebox{-0.5ex}{$\stackrel{>}{\scriptstyle{\sim}}$}
}\;}
\newcommand{\lsim}{\;\mbox{\raisebox{-0.5ex}{$\stackrel{<}{\scriptstyle{\sim}}$}
}\;}

\newcommand{\gb}{\mathcal{G}}

\newcommand{\M}{\mathcal{M}}

\begin{document}
\title{Baryogenesis via Gravitational Spontaneous Symmetry Breaking}

\author{Qiuyue Liang}
\email[Email: ]{qyliang@sas.upenn.edu}
\affiliation{Center for Particle Cosmology, Department of Physics and Astronomy, University of Pennsylvania 209 S. 33rd St., Philadelphia, PA 19104, USA}
\author{Jeremy Sakstein}
\email[Email: ]{sakstein@physics.upenn.edu}
\affiliation{Center for Particle Cosmology, Department of Physics and Astronomy, University of Pennsylvania 209 S. 33rd St., Philadelphia, PA 19104, USA}
\author{Mark Trodden}
\email[Email: ]{trodden@physics.upenn.edu}
\affiliation{Center for Particle Cosmology, Department of Physics and Astronomy, University of Pennsylvania 209 S. 33rd St., Philadelphia, PA 19104, USA}

\begin{abstract}
We study baryogenesis in effective field theories where a $\mathrm{U}(1)_{ B-L}$-charged scalar couples to gravity via curvature invariants. We analyze the general possibilities in such models, noting the relationships between some of them and existing models, such as Affleck-Dine baryogenesis. We then identify a novel mechanism in which $\mathrm{U}(1)_{ B-L}$ can be broken by couplings to the Gauss-Bonnet invariant during inflation and reheating. Using analytic methods, we demonstrate that this mechanism provides a new way to dynamically generate the net matter-anti-matter asymmetry observed today, and verify this numerically.
\end{abstract}
\maketitle

\section{Introduction}

There is more matter than anti-matter in the universe. This fact by itself is hardly surprising: baron and lepton number are accidental symmetries of the standard model that are broken by higher-dimensional operators and non-perturbative objects (sphalerons) \cite{Trodden:1998ym,Riotto:1999yt,Cline:2006ts,Dine:2003ax,Trodden:2004mj,Morrissey:2012db,Allahverdi:2012ju,Arbuzova:2018gsi,Racker:2014yfa,Cui:2015eba} so that the only conserved quantity is their difference, $B-L$. One would therefore expect some asymmetry in the universe, the exact amount of which can be quantified by the baryon to entropy ratio $n_{\rm B}/n_{\rm s}$, a quantity that is observed to be $\sim 10^{-10}$ \cite{Fields:2014uja,Cyburt:2015mya}. Whether or not the standard model can account for this depends on the magnitude of CP-violating effects and, crucially, on the nature of the electroweak phase transition. In particular, only a first-order transition can produce a sufficient asymmetry. The measurement of the Higgs mass at $125$ GeV (e.g. \cite{Aad:2015zhl}) implies that the transition is a crossover, and therefore some new physics beyond the standard model (BSM) is required to generate the observed asymmetry dynamically.

There is a profusion of BSM mechanisms and models that could generate such an asymmetry (see \cite{Chen:2007fv,Davidson:2008bu,Balazs:2014eba,Cline:2006ts} for reviews). Many BSM theories, including supersymmetric theories, include new scalar particles that are charged under $\mathrm{U}(1)_{\rm B-L}$ (this is a consequence of the fact that this combination is anomaly-free), and several baryogenesis mechanisms employ a breaking of $B-L$ to achieve a dynamical generation of the asymmetry, transferring it to the standard model at a later time, or through sphaleron processes. A particularly well-studied mechanism for breaking the symmetry in these scenarios is the Affleck-Dine mechanism \cite{Affleck:1984fy,Dine:1995kz,Dine:2003ax,Allahverdi:2012ju}. Here, the $\mathrm{U}(1)_{\rm B-L}$ can be broken during reheating due to a coupling between the scalar's kinetic energy and the inflaton. If the reheating is perturbative, i.e. the inflaton is oscillating about the minimum of its potential, then this coupling will dominate over any non-derivative couplings, and baryogenesis may proceed due to the symmetry breaking. However, the process of reheating is likely more complicated and non-linear than simple perturbative scenarios \cite{Allahverdi:2010xz,Amin:2014eta}, and this has motivated recent searches for symmetry breaking mechanisms that are less sensitive to the precise details of inflation and reheating, or decoupled from them completely. In particular, Ref.~\cite{Sakstein:2017lfm} presented a new model where a coupling of a scalar to dark matter or, equivalently, to the Ricci scalar, can provide the symmetry breaking needed without any reference to inflation whatsoever. Similarly, Ref.~\cite{Sakstein:2017nns} constructed a similar mechanism in which the symmetry is instead broken by a coupling to a new vector that spontaneously breaks Lorentz-invariance when it aligns with the cosmic rest frame.

The purpose of this paper is to systematically study the different mechanisms by which a $\mathrm{U}(1)$ symmetry can be broken by gravitational effects. To accomplish this, we consider the low energy effective field theory (EFT) of a $\mathrm{U}(1)$-charged scalar (that we will eventually identify with $B-L$) coupled to gravity via curvature tensors. We will also include potential couplings to the inflaton for illustrative purposes, and to make contact with the Affleck-Dine literature. In doing so, we will identify a new symmetry breaking mechanism that arises due to a coupling between the scalar and the Gauss-Bonnet (GB) invariant. This symmetry breaking exhibits novel features compared with the Affleck-Dine and scalar-dark matter scenarios, which we exploit to devise a new mechanism for baryogenesis: during inflation, the symmetry is broken and the field sits at its constant, symmetry-breaking minimum. After inflation, and during reheating, the GB term redshifts, which causes the symmetry breaking minimum to move closer to zero, and the symmetry is ultimately restored once this term has redshifted sufficiently. During this time, the motion of the angular field generates a net $B-L$ dynamically. After reheating, the symmetry is restored so that the field has a zero vacuum expectation value (VEV). The asymmetry is later transferred to the standard model via couplings of the scalar to the left-handed sector, for example via a neutrino portal. This coupling therefore allows for a baryogenesis scenario with some similarities to the Affleck-Dine mechanism, but for more general reheating (and preheating) mechanisms e.g. \cite{Heckman:2019dsj}.

The paper is organized as follows: in the next section we take an effective field theory approach to categorizing how gravitational effects can give rise to baryogenesis in the types of models discussed above. In section \ref{sec:BGBC} we then focus on a novel baryogenesis mechanism employing a coupling to the Gauss-Bonnet invariant, and study its implications before summarizing our results in the final section.

\section{Baryogenesis with Scalar-Gravity Couplings}
\label{sec:GBEFT}

The building blocks of our theory are a massless spin-2 graviton $g_{\mu\nu}$ and a new scalar $\phi$ which has charge $q$ under $\mathrm{U}(1)_{\rm B-L}$. In order to make contact with the Affleck-Dine mechanism we will also include an inflaton field $\Phi$, but this will play no role in the mechanism described in this work. The action is:
\begin{align}
S&=\int\dd^4 x\sqrt{-g}\left[\frac{\mpl^2}{2}R-\frac{1}{2}\nabla_\mu\Phi\nabla^\mu\Phi-\nabla_\mu\phi\nabla^\mu\phi^\star-m_\phi^2|\phi|^2-\lambda|\phi|^4\right.\nonumber\\
&\left.-c\frac{|\phi|^2}{\mathcal{M}^2}\nabla_\mu\Phi\nabla^\mu\Phi -V_{\rm inf}(\Phi)-V_{\rm int}(\Phi,|\phi|)-{\alpha}{|\phi|^2}R\pm\frac{|\phi|^2}{\mathcal{M}^2}\mathcal{G}\right],\label{eq:EFT}
\end{align}
where $\mathcal{G}=R^{\alpha\beta\mu\nu}R_{\alpha\beta\mu\nu}-4R_{\mu\nu}R^{\mu\nu}+R^2$ is the Gauss-Bonnet invariant. Some explanation is in order. Equation \eqref{eq:EFT} is the effective field theory for our new scalar coupled to gravity and the inflaton up to dimension-six operators, with cut-off denoted by $\mathcal{M}$. We have not included every possible coupling (e.g. inflaton-Gauss-Bonnet, $G^{\mu\nu}\partial_\mu\phi\partial_\nu\phi$, or $\Phi^2R$ for example) in the interest of compactness; these operators are either unimportant or do not qualitatively change the dynamics. The inflaton self-potential $V_{\rm inf}(\Phi)$ is responsible for driving inflation, while the interaction potential
$V_{\rm  int}(\Phi,|\phi|)$ represents other possible interactions with the scalar $\phi$ that are not forbidden by symmetries. They are included for completeness and play no role in what follows so we will not concern ourselves with their specific form. Finally, we must supplement equation \eqref{eq:EFT} with symmetry violating terms
\begin{equation}\label{eq:symmbreak}
S_{\textrm{symmetry violating}}=-\int\dd^4 x\sqrt{-g}\left[\frac{\varepsilon}{4}\phi^4+\frac{\varepsilon^\star}{4}{\phi^\star}^4\right].
\end{equation}
These operators are ubiquitous in any theory where there are several new scalar degrees of freedom \cite{Dine:1995kz}. For example, if the $\mathrm{U}(1)_{\rm B-L}$ charge $q$ and there is another heavy charged field $\chi$ with charge $4q$ the leading-order interaction is
\begin{equation}
S_{\phi\chi}=-\int\dd^4 x\sqrt{-g}\left[\frac{\phi^4\chi^*}{\bar{\mathcal{M}}}+\textrm{ h.c.}\right].
\end{equation}
If $\chi$ acquires a non-zero VEV then we can integrate it out to find precisely equation \eqref{eq:symmbreak} with
\begin{equation}
\varepsilon=4\frac{\langle\chi^\star\rangle}{\bar{\mathcal{M}}}.
\end{equation}
In this example, we expect ${\langle\chi^\star\rangle}\ll{\bar{\mathcal{M}}}$ so that $|\varepsilon|\ll1$.

At the end of inflation (and possibly during reheating), the inflaton is either absent or dominated by its kinetic energy i.e. $V_{\rm int}\sim V_{\rm inf}\approx 0$. In this case, equations \eqref{eq:EFT} and \eqref{eq:symmbreak} imply an effective potential for $\phi$:
\begin{equation}\label{eq:Veff}
V_{\rm eff}(\phi,\phi^\star)=\left(m_\phi^2+c\frac{\dot{\Phi}^2}{\mathcal{M}^2}+\alpha R\pm\frac{\mathcal{G}}{\mathcal{M}^2}\right)|\phi|^2+{\lambda}|\phi|^4+\varepsilon\frac{\phi^4}{4}+\varepsilon^\star\frac{{\phi^\star}^4}{4},
\end{equation}
where $\dot{\Phi}^2$ is the kinetic energy of the inflaton. We will see presently that the sign of the Gauss-Bonnet term depends on the equation of state, so we have included both signs for now in order to be general. Note also that the effective potential can either be symmetry preserving or symmetry breaking depending on the sign of the coefficient of the quadratic term. In particular, if this term is negative, the symmetry can be broken, which allows for the possibility of baryogenesis. By inspection, one can discern three possibilities for this. First, if $c<0$ then the kinetic coupling of $\phi$ to the inflaton can drive a symmetry breaking. This is how the symmetry is broken in the Affleck-Dine mechanism \cite{Affleck:1984fy,Dine:1995kz,Dine:2003ax}. (In this case one has $\dot\Phi ^2\sim H^2\mpl^2/\mathcal{M}^2$ so that the symmetry breaking term is proportional to $c(\mpl^2/\mathcal{M}^2)H^2$.)
 If $\alpha<0$ and there is some amount of non-relativistic matter present, then the symmetry can be broken because $R\sim \rho_{\rm m}/\mpl^2$. This is the mechanism of reference \cite{Sakstein:2017lfm}. Finally, if $\pm\mathcal{G}<0$---with the sign chosen depending on the sign of $\cal G$---then the symmetry can be broken by the Gauss-Bonnet invariant. This possibility has not been considered previously and so we will focus on this exclusively in the present work.\footnote{Note that the bounds on scalar-gravity couplings of this form from GW170817 \cite{Sakstein:2017xjx,Baker:2017hug,Creminelli:2017sry,Ezquiaga:2017ekz} do not apply to our models since the scalar is cosmologically subdominant throughout the entire universe's history \cite{Franchini:2019npi}.}

For a Friedmann-Robertson-Walker universe
\begin{equation}
\dd s^2 =-\dd t^2 +a^2(t)\left(\dd r^2+r^2\dd\Omega_{S^2}^2\right)
\end{equation}
dominated by a perfect fluid with equation of state $w$, the Gauss-Bonnet invariant is
\begin{equation}
\gb= \left\{
\begin{array}{cc}
-\frac{64(1+3w)}{27(1+w)^4t^4}\,\,\, &w\ne-1\vphantom{.},\nonumber\\
24H^4 \,\,\,&w=-1.
\end{array}
\right.
\label{eq:GBCOSMO}
\end{equation}
This implies that if we choose the negative sign in equation \eqref{eq:Veff} the symmetry is broken when $-1\le w<-1/3$ (i.e. whenever the universe is accelerating, including during an exact de Sitter phase). If instead we choose the positive sign, then the symmetry is broken if $w>-1/3$ (i.e. whenever the universe is decelerating). Therefore, the choice of sign is determined by the cosmological epoch during which one requires the symmetry to be broken. In order to decide this, it is instructive to consider the symmetry breaking minimum, which, using equation \eqref{eq:Veff}, is given by (ignoring the Ricci-coupling and the symmetry breaking terms for now)
\begin{equation}\label{eq:phiminn}
|\phi|_{\rm min}=\frac{1}{\sqrt{2\lambda}}\left(\left\vert\frac{\mathcal{G}}{\mathcal{M}^2}\right\vert-m_\phi^2\right)^{\frac{1}{2}}.
\end{equation}
During inflation (or to be more precise, in exact de Sitter space), the position of this minimum is constant, but on a general cosmological solution it is time-dependent and one has
\begin{equation}
|\phi|_{\rm min}=\frac{1}{\sqrt{2\lambda}}\left(\frac{64|(1+3w)|}{27(1+w)^4\mathcal{M}^2t^4}-m_\phi^2\right)^{\frac{1}{2}}.
\end{equation}
This is the first difference between Gauss-Bonnet-induced symmetry breaking and Affleck-Dine or dark-matter induced baryogenesis. In the latter two mechanisms, the symmetry-breaking term redshifts as $t^{-2}$, whereas in the former it redshifts as $t^{-4}$, allowing for a novel phenomenology. {Considering an epoch where $w$ can be treated as a constant,} the symmetry is restored at a time
\begin{equation}\label{eq:tbar}
\bar t^4=\frac{64|(1+3w)|}{27(1+w)^4\mathcal{M}^2m_\phi^2}.
\end{equation}
Now, since there is a strong time-dependence, it is not necessarily the case that the minimum is an attractor. Indeed, changing variables to
\begin{equation}
t=\bar te^{z};\quad \quad \xi=\frac{\phi(z)}{\phi_{\rm min}(z)},
\end{equation}
the equation of motion for the scalar at early times when $t\ll\bar t$ is
\begin{equation}\label{eq:xi}
\xi''-\frac{3+5w}{1+w}\xi'+2\frac{1+3w}{1+w}\xi+m_\phi^2\bar t^2e^{-2z}\xi\left(\xi^2-1\right)=0.
\end{equation}
The equivalent equation for Affleck-Dine or scalar-dark matter mechanisms has a fixed point near $\xi=1$ (the fixed point is not exactly at $\xi=1$ due to non-linear effects and cosmological time-dependence) \cite{Dine:1995kz}. In our case, setting $\xi'=\xi''=0$ one has
\begin{equation}
\xi^2=1-2\frac{e^{2z}}{m_\phi^2\bar t^2}\frac{1+3w}{1+w} \ ,
\end{equation}
so that on long enough time-scales (i.e. $t\gg\bar t$) the field will move away from the minimum exponentially. This is a new feature of the scalar-Gauss-Bonnet coupling. In practice, we are interested in the dynamics when $t\ll\bar t$ --- in fact, the approximations used to derive equation \eqref{eq:xi} break down at times $t\sim\bar t$ --- so this instability is not important. The important term for determining the stability in equation \eqref{eq:xi} is the second one, which represents a negative friction that will drive the field away from the minimum unless $-1\le w<-3/5$. This means that in order for baryogenesis to proceed in the absence of fine-tuning, any mechanism based on a scalar-Gauss-Bonnet coupling must necessarily occur during an epoch where the equation of state lies within this range. Clearly, this is not the case during the matter and radiation dominated eras. The patent epoch is reheating, which occurs at some time between the end of inflation, where the equation of state $w\approx-1$, and the start of the matter era, where $w=1/3$\footnote{One could imagine this process happening during inflation, but this would imply {isocurvature modes}, and could possibly alter the dynamics of inflation itself. It is also highly likely that any net $B-L$ generated would rapidly redshift unless one fine-tunes the parameters to have the requisite amount of $B-L$ produced very close to the end of inflation.}. The scenario we will therefore develop envisions that during reheating the equation of state changes from $w=-1$ to $w=1/3$ due to the production of radiation. During this time, the symmetry is broken and the minimum is an attractive fixed point, allowing for stable baryogenesis. We will now derive the dynamics of this mechanism in detail.

\section{Baryogenesis from a Gauss-Bonnet Coupling}
\label{sec:BGBC}

Having understood the dynamics of symmetry breaking due to a scalar-Gauss-Bonnet coupling, we are now in a position to describe a scenario for baryogenesis using this mechanism. Since such a scenario can only be viable during reheating, and when $-1\le w<-3/5$, we fix the sign of the scalar-Gauss-Bonnet coupling to be negative in equation \eqref{eq:Veff} so that the $\mathrm{U}(1)_{\rm B-L}$ can be broken spontaneously during this epoch (see the discussion below equation \eqref{eq:GBCOSMO}). The scenario is as follows.

\subsection{Inflation and Initial Conditions}

During inflation, the absolute value of the field $|\phi|$ sits at the (constant) minimum of the effective potential \eqref{eq:Veff}. Considering slow-roll inflation, which is the scenario favored by Planck observations \cite{Akrami:2018odb}, we can ignore the kinetic coupling to the inflaton, but one should not ignore the Ricci-coupling in a consistent effective field theory. The effective potential for this field is then
\begin{equation}
V(|\phi|)=\left(m_\phi^2+12\alpha H_I^2-24\frac{H_I^4}{\M^2}\right)|\phi|^2+\lambda|\phi|^4,
\end{equation}
where $H_I$ is the Hubble constant during inflation and $R=12H_I^2$. Provided that $H_I>(\alpha/2)^{1/2}\M$, the symmetry is broken during inflation and the field sits at the minimum, approximately given by
\begin{equation}
|\phi_I|\sim \sqrt{\frac{12}{\lambda}}\frac{H_I^2}{\M}.
\end{equation}
The mass at the minimum is of order $H_I^2/\M\gsim H_I$, so that the field is sufficiently heavy that it does not acquire isocurvature perturbations, nor does it impact the dynamics of the inflaton. Writing $\phi=|\phi|e^{i\theta}$, there is a small potential for the angular field $\theta$ due to the symmetry breaking terms
\begin{equation}
\Delta V(\theta)=\frac{\varepsilon_0}{2}|\phi|^4\cos(4\theta+\psi),
\end{equation}
where we have written $\varepsilon=\varepsilon_0\exp(i\psi)$. Recalling that the angular field is not canonically normalized, so that $\mathcal{L}\supset |\phi|^2(\nabla_\mu\theta)^2$, we infer that the mass for this field is
\begin{equation}\label{eq:mthetaI}
m_\theta^2\sim \varepsilon_0|\phi_I|^2\cos(4\theta_0+\psi)\sim \varepsilon_0\frac{H_I^2}{\M}\cos(4\theta_0+\psi),
\end{equation}
where $\theta_0$ is the initial value of $\theta$ during inflation. Now, since $\varepsilon_0\ll1$, we expect this mass to be sub-Hubble, so that the field is over-damped and is fixed to $\theta_0$. In fact, if this is not the case the field would minimize its potential during inflation and no net $B-L$ would be produced thereafter. This implies that $\theta$ acquires isocurvatuve perturbations \cite{Enqvist:1998pf,Enqvist:1999hv,Kawasaki:2001in}
\begin{equation}\label{eq:RRR}
\langle\delta\theta\rangle=\frac{H_I}{2\pi |\phi_I|}.
\end{equation}
Since the symmetry breaking terms are responsible for the generation of the net $B-L$, the amplitude of these perturbations is given by
\begin{equation}
S_{B\gamma}\equiv\frac{\delta \rho_B}{\rho_B}-\frac34\frac{\delta \rho_\gamma}{\rho_\gamma}=\frac{\delta n_B}{n_B},
\end{equation}
where the subscript $B$ denotes baryons and the subscript $\gamma$s denotes photons. Ultimately $n_B$ arises from $n_{B-L}$ after sphaleron reprocessing of the generated $B-L$. Therefore, we can estimate the amplitude of isocurvature perturbations by estimating $n_{B-L}$. This is generated by the motion of the angular field, so we have
\begin{equation*}
n_{B-L}\propto\partial_\theta\Delta V(\theta)\propto \sin(4\theta+\psi) \ ,
\end{equation*}
which yields ${\delta n_B}/{n_B}\sim 4\cot(4\theta_0+\psi)\delta\theta$. Using equation \eqref{eq:RRR}, and the isocurvature bounds from Planck \cite{Akrami:2018odb}, $S_{B\gamma}<3\times10^{-4}$, we have
\begin{equation}\label{eq:IB}
\frac{H_I}{|\phi_I|}=\frac{\pi}{2}\tan(4\theta_0+\psi)S_{B\gamma}\lsim 3\times10^{-4}.
\end{equation}
This imposes a bound on the initial conditions, and on the model parameters. We thus restrict to values where this is satisfied. It turns out this is not a very restrictive requirement, and we will verify later that it is satisfied by any reasonable model.

\subsection{Reheating and Baryogenesis}

When inflation ends, the equation of state increases from $-1$ through reheating, ultimately ending up at $1/3$, which signifies the beginning of the radiation epoch. During this phase, the effective potential is now given by
\begin{equation}
V(|\phi|)=\left(m_\phi^2+\alpha R_{\rm RH} + \frac{64(1+3w)}{27(1+w)^4\mathcal{M}^2t^4}\right)|\phi|^2+\lambda|\phi|^4,
\end{equation}
where $R_{\rm RH}$ denotes the Ricci scalar during reheating. To fully determine $R_{RH}$ would require us to specify the full (time-dependent) matter content during reheating, which would make the scenario highly model-dependent. In order to remain as agnostic as possible, we take $R_{\rm RH}\sim \rho_{\rm vac}/\mpl^2$, where $\rho_{\rm vac}$ is the vacuum energy density during reheating. Including only the contributions from the standard model, this is of order $\rho_{\rm EW}\sim200\textrm{ GeV}^4$ (assuming we reheat above the electroweak phase transition) so that the contribution of the Ricci coupling to the effective mass is of order $\sqrt{\rho_{\rm EW}}/\mpl^2\sim 10^{-14}$ GeV. This means we must take the mass of the field to be at least this large in what follows. It is possible that the vacuum energy is significantly larger if there are hidden-sector phase transitions in the early universe, so it is likely that the true value is somewhat larger than this. To parameterize our ignorance of this, we define
\begin{equation}
\bar{m}_\phi^2=m_\phi^2+\alpha R_{\rm RH},
\end{equation}
which we take to be constant. We require $m_\phi<\M$ in order to be consistent with the effective field theory (allowing $m_\phi>\M$ is tantamount to ignoring new light states that enter above the cut-off) but note that there is no such restriction on $\bar{m}_\phi$. With this in mind, the time-dependent minimum is given by
\begin{equation}
|\phi_{\rm min}(t)|=\frac{\bar{m}_\phi}{\sqrt{2\lambda}}\left[\left(\frac{\bar t}{t}\right)^4-1\right]^{\frac{1}{2}},
\end{equation}
where $\bar t$ is given by equation \eqref{eq:tbar}. The symmetry is restored at a time $t=\bar t$ and, for the sake of providing a concrete scenario, we assume that the equation of state crosses $w=-1/3$ at a time later than $\bar t$ so that the symmetry is restored due to the redshifting of the Gauss-Bonnet term but not due to the sign change that occurs when the equation of state exceeds $-1/3$.\footnote{A more stringent requirement is that the equation of state should exceed $-3/5$ at a time $t>\bar t$ so that the minimum is always a stable attractor. In practice, this distinction is negligible since the field begins near the minimum and tracks it for a long time in both cases.} The Hubble parameter when the symmetry is restored is
\begin{equation}\label{eq:Hbar}
\bar H = \frac{2}{3(1+w)\bar t}\sim \sqrt{\bar{m}_\phi\M}.
\end{equation}
In order to generate a net $B-L$, the angular field must begin to roll at around this time so that its mass $m_\theta\sim \bar H$. This gives the condition
\begin{equation}
\bar{m}_\theta^2=\frac{\Delta V_{\theta\theta}}{|\phi|_{\rm min}^2}\sim \varepsilon_0|\phi|_{\rm min}^2\sim \bar{H}^2,
\end{equation}
which allows us to fix the magnitude of the symmetry breaking terms
\begin{equation}\label{eq:eps}
\varepsilon_0\sim \frac{\bar H^2}{|\phi|_{\rm min}^2}\sim \lambda \frac{ \M}{ \bar{m}_\phi}.
\end{equation}
The motion of the angular field generates the $B-L$ asymmetry. In particular, the time-component of the conserved $\mathrm{U}(1)_{B-L}$ current, which corresponds to $n_{B-L}$, is
\begin{equation*}
n_{B-L}=j^0=2q\textrm{Im}(\phi^\star\partial^0\phi)=2q|\phi|^2\dot\theta \ ,
\end{equation*}
where $q$ is the $B-L$ charge of the scalar. One can see that angular motion is necessary to generate a net $B-L$. The equation of motion for the angular field can be expressed in the form
\begin{equation}\label{eq:dotnb}
\dot{n}_{B-L}+3Hn_{B-L}=2q\varepsilon_0|\phi|^4\sin(4\theta+\psi).
\end{equation}
Approximating $\dot{n}_{B-L}$ by $Hn_{B-L}$, using equation \eqref{eq:eps}, and ignoring order-unity factors, we have
\begin{equation}
n_{B-L}\approx\varepsilon_0\frac{|\phi|^4}{\bar H}\sin(4\theta+\psi)\sim |\phi|_{\rm min}^2\bar H \ .
\end{equation}
Since baryogenesis occurs during reheating, the entropy density around this time is
\begin{equation}
s\sim \frac{\mpl^2\bar H^2}{T_{\rm RH}}
\end{equation}
so that the ratio
\begin{equation}\label{eq:NBL}
\frac{n_{B-L}}{s}\sim 10^{-10}\left(\frac{10^{-5}}{\lambda}\right)\left(\frac{T_{\rm RH}}{10^{12}\textrm{ GeV}}\right)\left(\frac{\bar{m}_\phi}{10^{8}\textrm{ GeV}}\right)^{\frac32}\left(\frac{\M}{10^6\textrm{ GeV}}\right)^{-\frac12}.
\end{equation}
Equation \eqref{eq:NBL} is the main result of this section. It demonstrates that the mechanism that we have developed in this section can generate the observed matter-anti-matter asymmetry observed in the universe with suitable parameter choices. As it stands, this asymmetry is entirely stored in the new scalar $\phi$ but, provided that the scalar can decay to the left handed sector of the standard model, sphalerons will reprocess this into a net baryon and lepton asymmetry. The potential couplings of such scalars to the standard model were discussed at length in \cite{Sakstein:2017nns} and we have nothing new to add here. We simply remark that many such couplings are possible, and constraints from direct and indirect detection as well collider bounds are not yet at the level where such couplings are sufficiently forbidden.

\begin{figure}[t]
\includegraphics[width=0.55\textwidth]{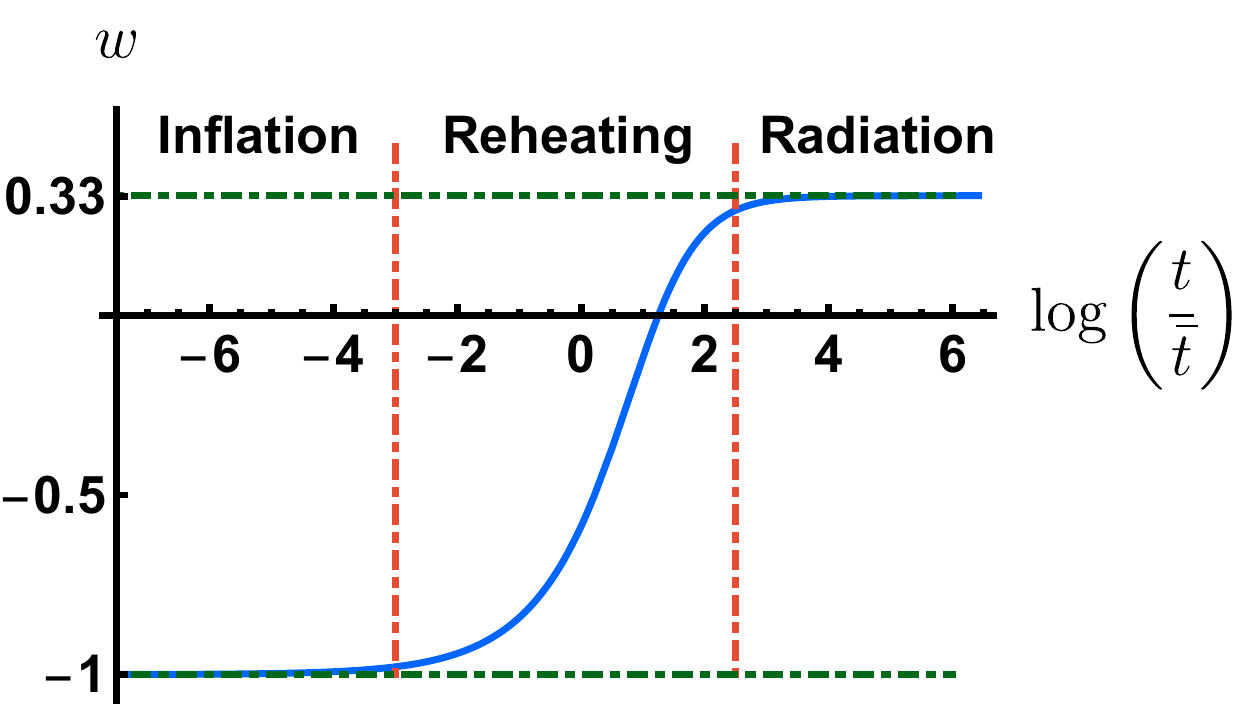}
\caption{The equation of state as a function of time that gives rise to the parameterization of the Hubble constant given in equation \eqref{eq:HPARA}. } \label{fig:eos}
\end{figure}

Some comments are in order. First, we have made several approximations. This was both for simplicity (e.g. ignoring order-unity factors, assuming a constant equation of state $w$ etc.) and to make the calculation analytically tractable. For example, we approximated $\dot{n}_{B-L}$ by $Hn_{B-L}$, and assumed that the evolution of the scale factor is given by the standard Friedmann equations without accounting for the backreaction of the Gauss-Bonnet term on the equation of motion of the metric. Said another way, we assumed that the  contribution of the scalar field to the evolution of the universe was subdominant throughout this entire process. This latter assumption is easy to verify. The contribution of the scalar potential to the Friedmann equation is
\begin{equation}
\Omega_\phi\sim \frac{\bar{m}_\phi^2|\phi|^2}{\mpl^2H^2}\sim \frac{\bar{m}_\phi^4}{\lambda\mpl^2\bar H^2}\sim\frac{m_\phi^3}{\lambda\mpl^2\mathcal{M}}\lsim10^{-12},
\end{equation}
where we have used equations \eqref{eq:phiminn}, \eqref{eq:Hbar}, and the the fiducial values needed to set every term in equation \eqref{eq:NBL} to unity. Similarly, the contribution from the Gauss-Bonnet coupling is
\begin{equation}
\Omega_{\mathcal{G}}\sim \frac{|\phi|^2}{\M^2}\frac{H^4+\dot{H}H^2}{H^2\mpl^2}\sim\frac{|\phi|^2}{\mpl^2}\frac{H^2}{\M^2}\lsim 10^{-12}.
\end{equation}
Evidently, the backreaction is indeed negligible. The other assumptions require numerical computations to verify. We do precisely this in the next section, where we solve the equations of motion numerically and verify equation \eqref{eq:NBL}.

\subsection{Numerical Examples}

\begin{figure}[ht]
\includegraphics[width=0.45\textwidth]{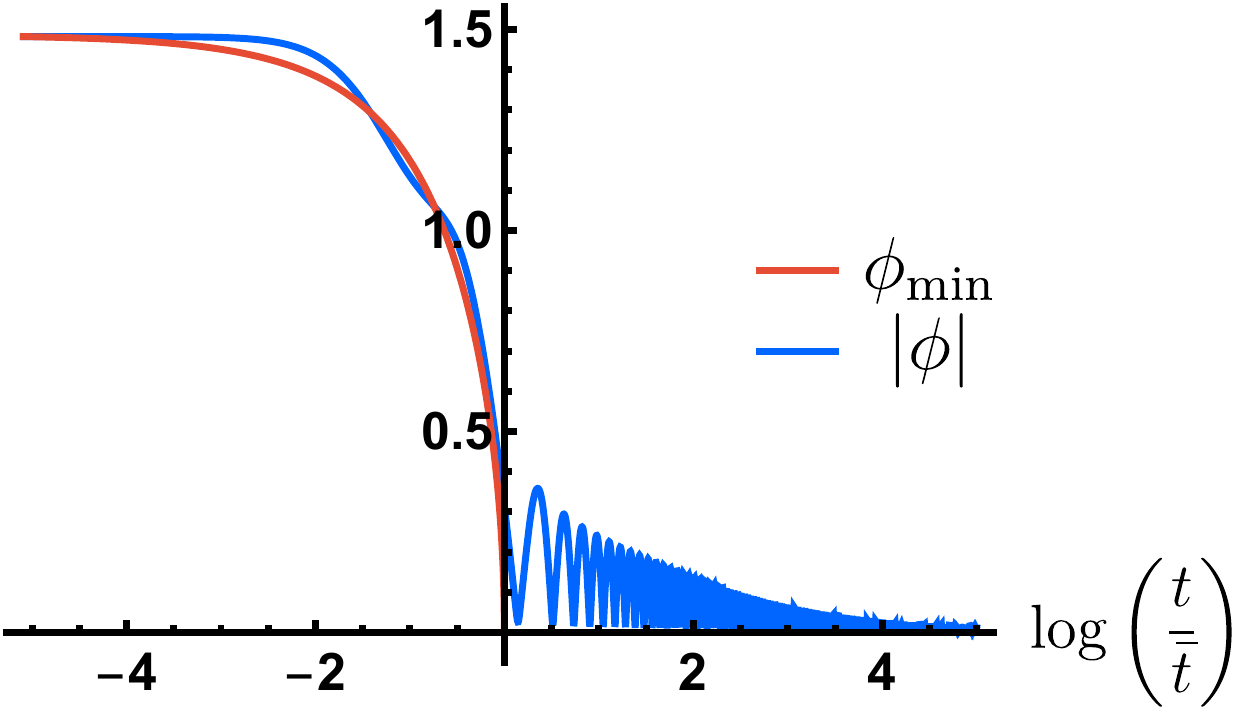}\;
\includegraphics[width=0.35\textwidth]{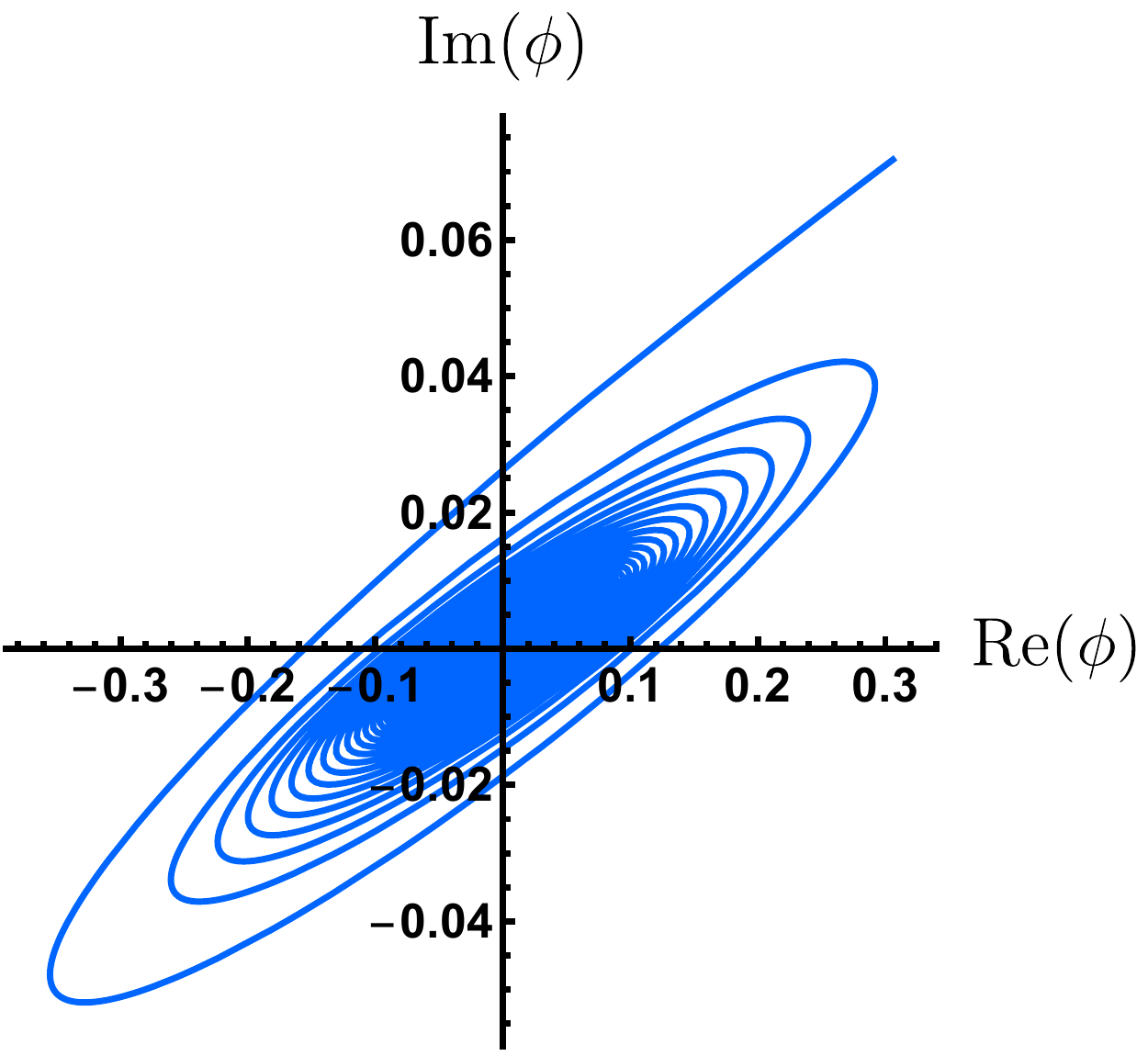}
\caption{\textbf{Left:} The evolution of $|\phi|_{\rm min}(t)$ calculated analytically (using the parameterization in equation \eqref{eq:HPARA}) and the results of the numerical integration for $|\phi|(t)$. \textbf{Right:} The evolution of real and imaginary parts of the scalar.} \label{fig:scalar}
\end{figure}

In this section we will verify numerically that baryogenesis during reheating from a scalar-Gauss-Bonnet coupling can indeed generate the net matter-anti-matter asymmetry observed in the universe as predicted analytically by equation \eqref{eq:NBL}, validating our approximations made in the previous sections. Since we have already established that the backreaction of the scalar on the Friedmann equations is negligible, we will only solve the equation of motion for the scalar
\begin{equation}\label{eq:SEOM}
\ddot{\phi}+3H\dot{\phi}+\frac{\partial V_{\rm eff}(\phi,\phi^\star)}{\partial \phi^\star}=0 \ ,
\end{equation}
with $V_{\rm eff}$ given by
\begin{equation}
V_{\rm eff}(\phi,\phi^\star)=\left(\bar{m}^2_\phi-24\frac{H^4+H^2\dot{H}}{\M^2}\right)|\phi|^2+\lambda|\phi|^4+\frac{\varepsilon}{4}\phi^4+\frac{\varepsilon^\star}{4}{\phi^\star}^4.
\end{equation}
i.e. we are now including the symmetry breaking terms and the effective mass $\bar{m}$ in its definition. In order to be agnostic about the details of reheating, we parameterize the Hubble constant during reheating as
\begin{equation}\label{eq:HPARA}
H(t)=\frac{H_I}{\left[1+4(H_I t)^2\right]^{\frac12}} \ ,
\end{equation}
so that it interpolates between a constant value $H_I$ during inflation and tends to $1/(2t)$ at late times, corresponding to a radiation dominated universe. The intermediate epoch is presumed to encapsulate reheating. Note that there is no period where $H=2/3t$ (i.e. $w=0$) so this parameterization does not include perturbative reheating, which is the preferred epoch for Affleck-Dine baryogenesis. As discussed above, our mechanism would not work during perturbative reheating because the minimum would not be a stable attractor. The equation of state as a function of time for this parametrization is shown in figure \ref{fig:eos}. As a consistency check, we integrated the equations with several alternate parameterizations; all gave qualitatively similar results.

\begin{figure}[ht]
\includegraphics[width=0.7\textwidth]{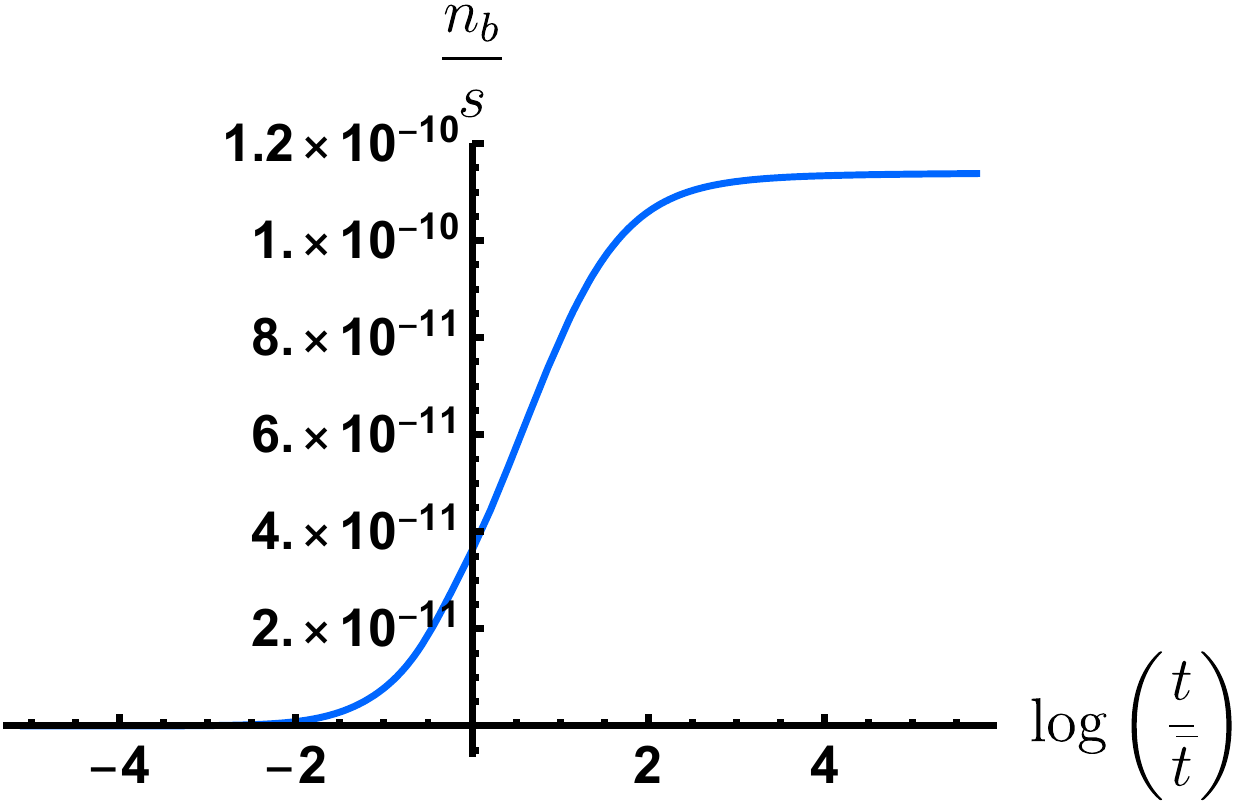}
\caption{The ratio $n_{B-L}/s$ as a function of time. } \label{fig:nbs}
\end{figure}

The results of our integrations are shown in figures \ref{fig:scalar} and \ref{fig:nbs}. In all cases we used the fiducial values $\bar{m}_\phi=7\times10^{8}$ GeV, $\M=10^6$ GeV, $H_I=1.6\times10^{7}$ GeV, $T_{\rm RH}=2.1\times 10^{12}$ GeV, $\lambda=3.3\times10^{-4}$, $\varepsilon_0 = 3.2\times 10^{-7}$
and checked that qualitatively similar results are obtained for other choices. The isocurvature bound given in equation \eqref{eq:IB} can also be satisfied for this parameter choice. The evolution of the scalar is shown in figure \ref{fig:scalar}. The left panel shows the evolution of the scalar compared with the theoretical prediction for the evolution of the minimum. One can see that the field adiabatically tracks the minimum until the symmetry is restored. The right panel shows the real and imaginary parts of the scalar, and the angular motion that generates the $B-L$ asymmetry is evident. Finally, we plot the ratio $n_{B-L}/s$ in figure \ref{fig:nbs}. A value of order $10^{-10}$ is reached asymptotically, confirming our approximations in the previous sections, and their ultimate analytical prediction given in equation \eqref{eq:NBL}. We have found that changing the initial conditions and the parameters yields qualitatively similar behavior.

\section{Discussion and Conclusions}

In this work we have studied the effective field theory of a $\mathrm{U}(1)$ scalar {field} coupled to both gravity and the inflaton. This theory contains several natural mechanisms for spontaneously breaking the $\mathrm{U}(1)$ symmetry, including the Affleck-Dine mechanism \cite{Affleck:1984fy,Dine:1995kz,Dine:2003ax}, the matter-induced breaking of \cite{Sakstein:2017lfm}, and a novel mechanism where a coupling to the Gauss-Bonnet invariant can induce a tachyonic instability for the scalar around the symmetry-preserving point in field space. Identifying the charge with $B-L$, we have then developed a novel scenario for baryogenesis that employs this breaking mechanism.

The Gauss-Bonnet term has some unique properties that give rise to several novel features compared with other scenarios that make use of cosmological dynamics to realize an inverse symmetry breaking transition. First, the Gauss-Bonnet term has different signs depending on the equation of state, with the threshold given by $w=-1/3$. This means that the symmetry can either be broken during reheating and inflation, or during the standard big bang cosmological history, but not both. Second, the contribution to the effective mass of the scalar redshifts like $t^{-4}$ rather than $t^{-2}$ as is the case for Affleck-Dine and matter-induced baryogenesis. We have shown that this causes the (time-dependent) minimum to be an attractor when $-1<w<-3/5$ and an unstable fixed point otherwise, signaling that any stable baryogenesis mechanism must happen during a cosmological epoch dominated by a component with an equation of state in this range, inflation and reheating being the patent epochs.

Using the above facts, we have constructed the following scenario: during inflation, the symmetry is broken and the radial part of the scalar minimizes its effective potential (the position of the minimum is constant), while the angular field is frozen due to Hubble damping. After reheating begins, the minimum of the effective potential begins to move towards zero as the Gauss-Bonnet term stars to redshift. At this time, the angular field begins to roll down its potential, which arises due to small $\mathrm{U}(1)_{B-L}$-breaking terms in the action, generating a net $B-L$. Finally, the symmetry is restored when the Gauss-Bonnet term has redshifted sufficiently, and the radial part of the field rapidly returns to zero. We have analytically predicted the amount of $B-L$ produced, demonstrating that the net matter-anti-matter asymmetry observed in the universe can be accounted for using reasonable parameter choices. We have also verified our analytic predictions using numerical computations.

Of course, this is just the first part of the story. In order to fully account for the observed asymmetry the $B-L$ must be transferred into the left-handed sector of the standard model. This is not particularly difficult and there are several portals to the standard model that are allowed by current constraints \cite{Sakstein:2017lfm}. Any of these are sufficient for our purposes.

An important question is whether these theories can be tested observationally. Interestingly, theories of this kind exhibit a phenomenon known as \emph{spontaneous scalarization} \cite{Silva:2017uqg,Silva:2018qhn,Macedo:2019sem,Doneva:2017bvd,Doneva:2017duq,Minamitsuji:2018xde} whereby black holes and neutron stars can suddenly acquire a large scalar charge (or, equivalently, hair) due to quadratic scalar-Gauss-Bonnet coupling. Indeed, it is precisely the same tachyonic instability we identified here that underlies this mechanism. So far, all of the studies of spontaneous scalarization have used real fields, so it would be interesting to generalize these to complex fields. In particular, Ref.~\cite{Macedo:2019sem} studied an identical EFT to equation \eqref{eq:EFT} (without the inflaton coupling) with the exception that the scalar field there was real. The cut-off for theories that scalarize solar mass black holes is $\M\sim 10^{-22}$ GeV ($\sim 35$ km) \cite{Macedo:2019sem}, several orders of magnitude smaller than the cut-off for our baryogenesis mechanism. It would be interesting to explore which type of objects could become scalarized in our theory.  Recently, reference \cite{Andreou:2019ikc} studied the most general subset of Horndeski theories that can give rise to a tachyonic mass for a real scalar field. It would be interesting to generalize this to complex scalars, and determine whether any new baryogenesis mechanisms exist.

\acknowledgements
JS is supported by the Center for Particle Cosmology at the University of Pennsylvania. The work of MT is supported in part by US Department of Energy (HEP) Award DE-SC0013528, and by NASA ATP grant NNH17ZDA001N.

\bibliography{ref}

\end{document}